\documentclass[aps,prb,twocolumn,superscriptaddress]{revtex4-2}

\usepackage{graphicx}
\usepackage{color}

\begin{document}


\title{Tunable Microwave Single-photon Source Based on Transmon Qubit with High Efficiency}


\author{Yu Zhou}
\email[]{yu.zhou@riken.jp}
\affiliation{Department of Physics, Tokyo University of Science, 1-3 Kagurazaka, Shinjuku-ku, Tokyo 162-8601,Japan}
\affiliation{Center for Emergent Matter Science, RIKEN, 2-1 Hirosawa, Wako, Saitama 351-0198, Japan}

\author{Zhihui Peng}
\affiliation{Key Laboratory of Low-Dimensional Quantum Structures and Quantum Control of Ministry of Education,	Department of Physics and Synergetic Innovation Center for Quantum Effects and Applications, Hunan Normal University, Changsha 410081, China}
\affiliation{Center for Emergent Matter Science, RIKEN, 2-1 Hirosawa, Wako, Saitama 351-0198, Japan}

\author{Yuta Horiuchi}
\affiliation{Department of Physics, Tokyo University of Science, 1-3 Kagurazaka, Shinjuku-ku, Tokyo 162-8601,Japan}

\author{O.V. Astafiev}
\affiliation{Skolkovo Institute of Science and Technology, Moscow 143026, Russia}
\affiliation{Physics Department, Royal Holloway, University of London, Egham, Surrey TW20 OEX, UK}
\affiliation{National Physical Laboratory, Teddington TW11 OLW, UK}
\affiliation{Moscow Institute of Physics and Technology, Dolgoprudny 141700, Russia}

\author{J.S. Tsai}
\email[]{tsai@riken.jp}
\affiliation{Department of Physics, Tokyo University of Science, 1-3 Kagurazaka, Shinjuku-ku, Tokyo 162-8601,Japan}
\affiliation{Center for Emergent Matter Science, RIKEN, 2-1 Hirosawa, Wako, Saitama 351-0198, Japan}


\date{\today}

\begin{abstract}
Single-photon sources are of great interest because they are key elements in different promising applications of quantum technologies. Here we demonstrate a highly efficient tunable on-demand microwave single-photon source based on a transmon qubit with the intrinsic emission efficiency above 98$\%$. The high efficiency ensures a negligible pure dephasing rate and the necessary condition for generation of indistinguishable photons. We provide an extended discussion and analysis of the efficiency of the photon generation. To further experimentally confirm the single-photon property of the source, correlation functions of the emission field are also measured using linear detectors with a GPU-enhanced signal processing technique. Our results experimentally demonstrate that frequency tunability and negligible pure dephasing rate can be achieved simultaneously and show that such a tunable single-photon source can be good for various practical applications in quantum communication, simulations and information processing in the microwave regime.

\end{abstract}


\maketitle

\section{\label{sec:intro}Introduction}
 Controllable single photons are an important tool to study fundamental quantum mechanics and also for practical applications in quantum communication \cite{kimble_quantum_2008}, sensing \cite{degen_quantum_2017}, simulations \cite{georgescu_quantum_2014} and computing \cite{knill_scheme_2001,kok_linear_2007}. Single-photon sources thus have been extensively studied in optics \cite{lounis_single-photon_2005,eisaman_invited_2011} and a great progress has been achieved \cite{ding_demand_2016,somaschi_near-optimal_2016,senellart_high-performance_2017,schweickert_-demand_2018}. The single-photon sources based on superconducting circuits \cite{you_superconducting_2005,devoret_superconducting_2013,gu_microwave_2017} in the microwave regime have also attracted great interest, having a unique property -- the easily achievable strong interaction with electromagnetic waves. This property allows to reach high efficiency in generating and detecting microwave photons. There have already been some implementations of single-photon sources, which are based on cavity QED systems \cite{houck_generating_2007,bozyigit_antibunching_2011,bozyigit_correlation_2011,lang_correlations_2013,pechal_microwave-controlled_2014}. Instead of confining the photons with a fixed cavity mode, recently several single-photon sources have also been demonstrated by strong coupling to one-dimensional (1D) continuum \cite{lindkvist_scattering_2014,roy_colloquium:_2017} and generating tunable single photons, using either flux qubits \cite{peng_tuneable_2016} or transmon qubits \cite{pechal_superconducting_2016,forn-diaz_-demand_2017,gasparinetti_correlations_2017}. 
 
However, for many practical applications, such as boson sampling \cite{aaronson_computational_2013}, the photons must be indistinguishable, which means that the pure dephasing should be suppressed \cite{santori_indistinguishable_2002,bylander_interference_2003,senellart_high-performance_2017}. For single-photon sources based on cavity QED systems, it will be fulfilled naturally due to the coupling to a fixed cavity mode. While for tunable ones, it can be achieved, when nearly perfect coupling to the 1D continuum is realized \cite{astafiev_resonance_2010}. To achieve this goal, we use the transmon qubit for its simplicity and better intrinsic coherence compared with flux qubits in Ref. \cite{peng_tuneable_2016}. By careful engineering the system, here we demonstrate a high quality tunable on-demand microwave single-photon source based on a transmon qubit with the intrinsic emission efficiency above $98\%$, which not only means the nearly perfect collecting efficiency of emitted photons but also, more importantly, shows the negligible pure dephasing rate experimentally. In earlier experiments with transmon qubits demonstrated in Refs. \cite{pechal_superconducting_2016,forn-diaz_-demand_2017}, the crucial pure dephasing rate is either not explicitly shown or not negligible. A systematic study of the single-photon source furthermore demonstrates the dynamics of emission field and the correlation function measurements with a GPU-enhanced signal processing technique, which confirms the single-photon emission. The theoretical numerical calculations using the Lindblad master equation with time-dependent Hamiltonian agree very well with the experimental results. Our results show that the frequency tunability and negligible pure dephasing rate can be achieved simultaneously in experiment. A further analysis of the efficiency indicates that such a tunable single-photon source can be a good source for various practical applications in quantum optics and quantum information in the microwave regime.
\section{\label{sec:dev}Device and experiment setup}
\begin{figure}
	\includegraphics[width=0.45\textwidth]{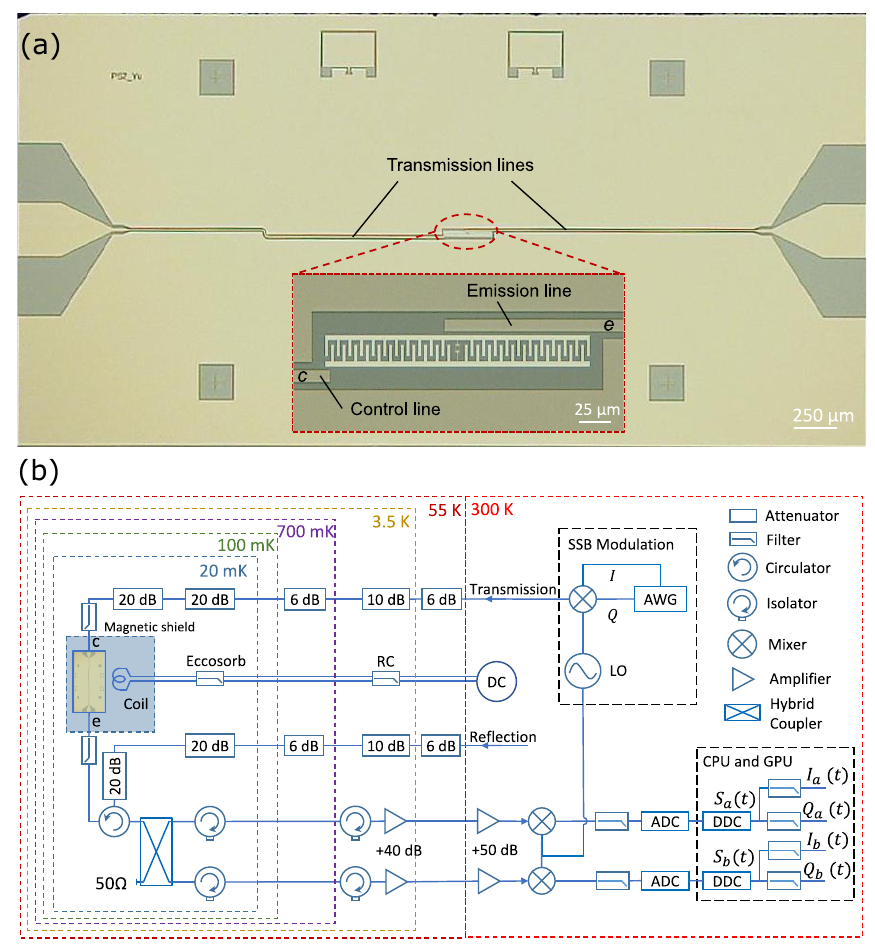}
	\caption{\label{fig1}(a) Optical image of the sample and inset shows the magnified structure of the single-photon source with a transmon qubit. (b) Schematic diagram of the cryogenic and room-temperature experimental setups for both spectrum and time-domain measurements.}
\end{figure}

Our single-photon source, see Fig.~\ref{fig1}(a), consists of a transmon qubit \cite{koch_charge-insensitive_2007} capacitively coupled to two open-ended 1D coplanar-waveguide transmission lines: one is weakly coupled to the transmon qubit to control its states (control line) and the other is strongly coupled for the photon emission (emission line), similar to Ref. \cite{peng_tuneable_2016}. Here the control and emission lines are coupled through a capacitance network, which includes the shunt capacitor of the transmon qubit and capacitances from the qubit electrodes to the ground. The effective attenuation is estimated using a EM simulator to be more than 80~dB at 7~GHz and the measured attenuation in off-resonance is more than 50 dB with the probe power of -140 dBm. 

The sample is fabricated using a standard fabrication technique for superconducting quantum circuits. The transmission line is made of the 50 nm thick Nb film on an undoped sillicon wafer. The qubit consisting of a dc-SQUID is fabricated with a standard Al/AlOx/Al shadow evaporation technique using an electron beam evaporation system. From the measured spectrum, the Josephson energy in maximum is $E_J^{max}/h\approx16.8$ GHz and the charging energy $E_C/h\approx415$ MHz, where $E_C = e^2/2C_q$ with an effective qubit capacitance $C_q\approx$~47~fF. The qubit energy is controlled by an external magnetic field. 

To study the single-photon source, the sample is cooled down to a temperature of about 20~mK in a dilution refrigerator with a microwave circuit shown in Fig.~\ref{fig1}(b). The sample is screened against external magnetic fields by two-layer $\mu$-metal and one-layer Al shield (see the area of magnetic shield in Fig.~\ref{fig1}(b)). Here the signal from the control line is strongly attenuated at different temperature stages of our dilution refrigerator to minimize the excitation of the transmon qubit by the room-temperature black-body radiation and an additional 20 dB attenuator (totally 40 dB attenuation) in the mixing chamber is to further suppress the thermal excitations. Two high rejection low-pass filters with cut-off frequency at 8 GHz are placed at both control and emission ports of the sample to additionally protect the transmon qubit from the high-frequency radiation noise. A 4-8 GHz circulator placed in the emission line allows to measure the reflection from the emission line, which can be used to characterize the coupling (and emission) efficiency of the single-photon source \cite{peng_tuneable_2016}. A 2-8 GHz hybrid coupler is placed at the output port working as a beam splitter in the HBT type setup \cite{brown_correlation_1956} to show the dynamics of the emission and also the correlation functions of the emitted radiation. An idle input port of the coupler has been terminated by a 50 $\Omega$ terminator. Additional isolators situated at the mixing chamber (MC) stage are to protect the source from the back-action of cryogenic amplifiers installed at 4K stage. The total gain of the output line is $\sim$ 90 dB, including the room-temperature (RT) amplifiers. A dc bias line with an RC filter (at 4K) and an Eccosorb low-pass filter (at MC) is used for the global flux bias to tune the transition frequency of the transmon qubit.

For the time-domain and correlation function measurements, we use the single-sideband (SSB) modulation technique with a single microwave source to stabilize the phase in the long-lasting experiment. The single-photon emission after linear amplification is down-converted to 25 MHz and digitized with an ADC at a sampling rate of 250 MS/s, then further processed by a CPU with a GPU-enhanced signal processing technique to extract the quadrature amplitude $S_{a/b}(t)$ and calculate correlation functions efficiently. 

Note, we use the linear detectors to carry out the correlation function measurements \cite{da_silva_schemes_2010,bozyigit_antibunching_2011,lang_correlations_2013}. Even though several microwave single-photon detectors have already been recently demonstrated \cite{inomata_single_2016,kono_quantum_2018,besse_single-shot_2018}, they are still far from practical applications in real-time microwave single-photon detection. So using linear detectors to carry out the correlation function measurement is so far a more practical and general approach.      
\section{\label{sec:char}Spectrum and Emission Efficiency}
Similar to Ref. \cite{peng_tuneable_2016}, firstly, we characterize our single-photon source by measuring the transmission from control line to emission line using a vector network analyzer (VNA). The transmission is strongly enhanced, when the drive signal at $\omega_d$ is in resonance with the transmon transition frequency $\omega_{01}$, which is a result of emission from the excited transmon qubit to the emission line under continuous microwave drive. As shown in Fig.~\ref{fig2}(a), the single-photon source can be tuned in the range from 4~GHz to 7~GHz. 

In the observed spectrum corresponding to the system resonance frequency $\omega_{01}$ (transition between $|0\rangle$ and $|1\rangle$ states), there are two avoided-crossings due to the coupling to two-level system (TLS) defects. The large offset in flux bias is caused by the residual magnetism in cables inside the magnetic shield which have been replaced in the later experiments. The linewidth ($-3$~dB in amplitude) at the sweet-point with $\omega_{m}/2\pi$ ~=~7.062~GHz (maximal $\omega_{01}$) is $\Delta\omega/2\pi\approx7$ MHz.      

Next, we characterize the efficiency of the emission from the transmon qubit to the emission line. The efficiency $\eta = \Gamma_1^e/\Gamma_1$ can be defined as the ratio of the emission rate $\Gamma_1^e$ over the total relaxation rate $\Gamma_1 = \Gamma_1^e + \Gamma_1^c + \Gamma_1^n$, where $\Gamma_1^c$ is the relaxation rate through the emission into the control line and $\Gamma_1^n$ is the non-radiative relaxation rate. We simulated our metallic structure and found from the capacitance network that the ratio of the relaxation rates to the lines $\Gamma_1^c/\Gamma_1^e$ is $(1.0\pm0.3)\times 10^{-2}$, taking into account the accuracy of derived capacitances. This means that ideally about 99\% of radiation can be emitted to the emission line. 

As described in Refs. \cite{peng_tuneable_2016,astafiev_resonance_2010}, the emission amplitude of coherent radiation is determined by the expectation value of the qubit annihilation operator $\langle\sigma^-\rangle$. 
By solving the master equation for the two-level system under continuous drive, we find $\langle\sigma^-\rangle=-i\frac{\Omega}{2\Gamma_2}\frac{1-i\delta\omega/\Gamma_2}{1+(\delta\omega/\Gamma_2)^2+\Omega^2/(\Gamma_1\Gamma_2)}$. Here $\delta\omega=\omega_d-\omega_{01}$ is the detuning of the drive, $\Omega$ is the Rabi frequency and $\Gamma_2={\Gamma_1}/2+\gamma$ is the dephasing rate, which includes pure dephasing rate $\gamma$. We can further find the reflection in the emission line $r_e$ as
\begin{equation}
r_e=1-\frac{\Gamma_1^e}{\Gamma_2}\cdot\frac{1-i\delta\omega/\Gamma_2}{1+(\delta\omega/\Gamma_2)^2+\Omega^2/(\Gamma_1\Gamma_2)}~.
\label{equ:1}
\end{equation} 
At the weak driving limit $\Omega\ll(\Gamma_1,\Gamma_2)$, it is simplified to 
\begin{equation}
r_e\approx 1-\frac{\Gamma_1^e}{\Gamma_2}\cdot\frac{1}{1+i\delta\omega/\Gamma_2}~.
\label{equ:2}
\end{equation}
and represents a circle with a radius of $\Gamma_1^e/2\Gamma_2$. Importantly, in the ideal case with $\Gamma_1^c = \Gamma_1^n = \gamma = 0$, the radius becomes equal to one and $\eta=1$.

As shown in Fig.~\ref{fig2}(b), we measure the reflection $r_e$ from the atom in the emission line (at the sweet point) with the probe power varied from a weak drive at $W = -146$ dBm up to a strong drive at $-116$ dBm. The data is normalized to its background measured when the qubit is tuned far away and the solid lines are the fitting results using Eq.~(\ref{equ:1}). The accuracy of normalization procedure is subjected to the uncertainty due to uncontrolled small amplitude and phase offset between two separately measured signal and background traces using VNA. We estimate that the contribution of such uncertainty to the fluctuations of final fitted efficiency could be $\pm0.01$, which is consistent with the experimental observation. Fitting the experimental curves using Eq.~(\ref{equ:1}) with all power traces, we found $\eta\ge\eta'=\Gamma_1^e/2\Gamma_2=0.991\pm0.011$ in the limit of $\Omega \rightarrow 0$. Here the error mainly comes from the uncertainty of normalization and the value can not exceed one. 
The efficiency value is consistent with our preliminary estimates and also confirms that there are no obvious thermal excitations, which would reduce the reflection due to the thermal population. Note here, the high emission efficiency indicates that nearly all photons are emitted into the emission line and the relaxation rates $\Gamma_1^c$ and $\Gamma_1^n$ together with the possible pure dephasing $\gamma$ are very weak. In that conditions, the linewidth is defined by the relaxation into the emission line $\Gamma_1^e/2\pi \approx$~7~MHz and the relaxation time $T_1^e = 1/\Gamma_1^e \approx$~23~ns. Such high emission efficiency also indicates that the qubit intrinsic (non-radiative) relaxation time is much higher than 2.3~ $\mu$s. For example, if the qubit intrinsic (non-radiative) relaxation time is 10~$\mu$s, it will reduce the efficiency by 23~ns$/10$~$\mu$s~$\approx 2\times10^{-3}$. Such a relaxation time or even higher is reasonable for the transmon qubit on silicon wafer with the currently used fabrication techniques. 

\section{\label{sec:eff}Limiting factors in the photon generation efficiency}
Here we briefly discuss some limitations on the maximal achievable photon generation efficiency. The efficiency is affected by several factors and the coupling efficiency discussed above is important but only one of them. In order to reach $(1-\alpha_p)$ (where $\alpha_p \ll 1$) efficiency of preparing the excited state $|1\rangle$, one needs to drive the qubit within time $\Delta t$, satisfying the condition $\Gamma_1^e\Delta t\approx\alpha_p$ and the required driving amplitude is $\Omega \approx \pi/\Delta t=\pi\Gamma_1^e/\alpha_p$. We also assume that the ratio of the couplings to two lines is $\alpha_c^2=\Gamma_1^c/\Gamma_1^e$. The corresponding applied power to the control line is $W=\hbar\omega\Omega^2/\Gamma_1^c=\hbar\omega \frac{\pi^2\Gamma_1^e}{\alpha_p^2\alpha_c^2}$. 
There is another obvious limitation on the minimal time $\Delta t_m$ allowed to prepare the excited state $|1\rangle$ by either the bandwidth of the equipment or the anharmonicity of the qubit, thus $\alpha_p\ge\Gamma_1^e\Delta t_m$. Obviously the coupling efficiency discussed above can be presented as $\eta = 1/(1+\alpha_c^2+\Gamma_1^n/\Gamma_1^e)$. If we assume realistic parameters $\alpha_c^2=0.01$, $\alpha_p=0.01$ and $\Delta t_m=2$~ns, then one should make $\Gamma_1^e\le(0.2~{\mu s})^{-1}$ and $\Gamma_1^n \ll \Gamma_1^c=\alpha_c^2\Gamma_1^e=(20~{\mu s})^{-1}$ to realize $\eta\approx0.99$, which can be achieved with the typical high quality qubit lifetime and proper engineering of the capacitance network. The total single-photon generation efficiency is then $(1-\alpha_c^2)(1-\alpha_p)\approx0.98$, which includes both state preparation and emission efficiency. 
Substituting the parameters, we find the necessary power within the pulse to generate the photons, which is $W \approx 2\times10^{-10}$~Watt at 7~GHz, however the total energy is only $W \Delta t_m\approx 4\times10^{-19}$~J. To further improve the efficiency (decrease of $\alpha_p$ and $\alpha_c^2$), the power still can be increased by about 4 -- 6 orders.  
Another factor to be taken into account is a direct leakage $\alpha_l=\frac{W\beta}{\hbar\omega\Gamma_1^e}=\frac{\pi^2\beta}{\alpha_p^2\alpha_c^2}$ of the coherent radiation from the control to the emission line due to the possible small stray capacitive coupling $\beta$ between two lines. The factor is very weak, less than 0.01, in our current device. 

Compared with our previous work \cite{peng_tuneable_2016}, where the emission efficiency is 0.79 with a linewidth of 20 MHz, here we achieve a much higher emission efficiency together with a much narrower linewidth of 7~MHz. This results in the higher excited state preparation efficiency of $1-\alpha_p\approx0.87$, with effective $\Delta t\approx3$~ns and $\Gamma_1^e\approx (0.023~\mu s)^{-1}$ in the experiment. The high emission efficiency achieved in this work is a result of the properly chosen device geometry and the much longer intrinsic coherence time of our transmon qubit. 

\begin{figure*}
	\includegraphics[width=0.9\textwidth]{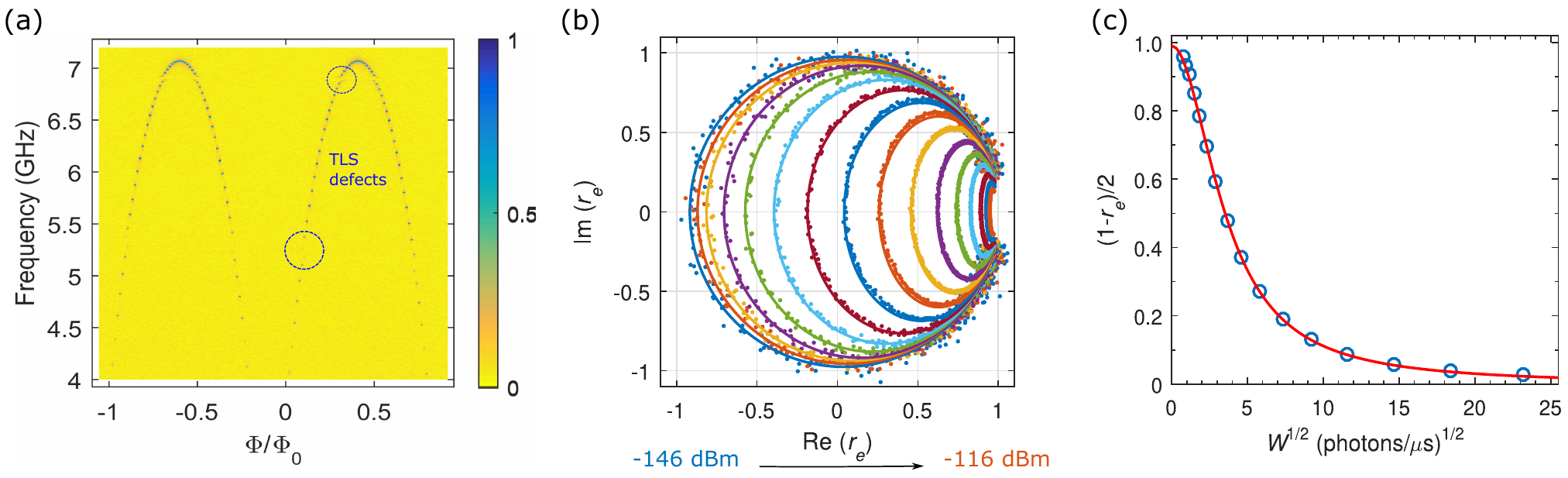}
	\caption{\label{fig2}(a) Normalized transmission spectrum $|t_{ce}/t_{max}|$ vs flux bias. The large offset in flux bias is caused by the residue magnetism in cables inside the magnetic shield. There are also two avoided-crossings because of the coupling to TLS defects. (b) Reflection $r_e$ at emission line when transmon is bias at the sweet-point. The experimental data (dots) are normalized to the background when $\omega_{01}$ is tuned far away from the sweet-point. The plot is in real and imaginary coordinates at the probe power from $-146$ dBm ($\sim0.5$ photons/$\mu$s) to $-116$ dBm ($\sim53.7$ photons/$\mu$s) with 2 dBm/step. The solid lines are the fitting results using Eq.~(\ref{equ:1}). (c) Plot of $(1-r_e)/2$ vs power $W$ when $\delta\omega = 0$. The dots are data extracted from the fitting results of (b) and the solid red line is $(1-r_e)/2=A/(1+kW)$, with $A = 0.991$ and a fitted factor $k$.}
\end{figure*}

\begin{figure}
	\includegraphics[width=0.45\textwidth]{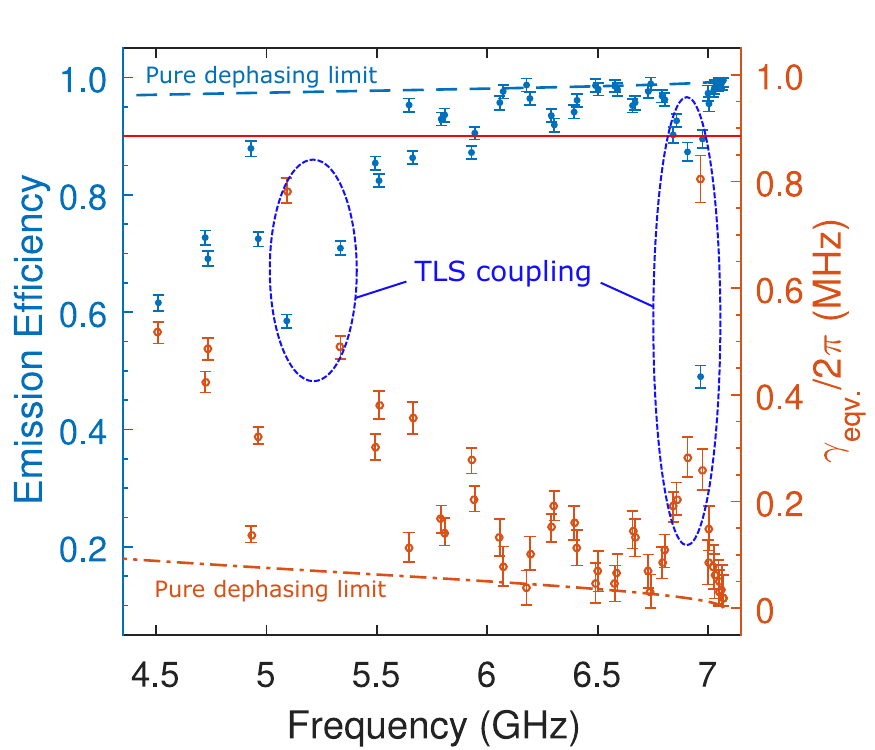}
	\caption{\label{fig3} Left $y$-axis: derived emission efficiency $\eta'$ over a wide range. The red line shows the position where efficiency is 90\%. The abnormal drop of $\eta'$ near the flux bias sweet-point is mainly caused by the coupling to TLS defect (see spectrum in Fig.~\ref{fig2}(a)). The dashed line shows maximum $\eta'$ limited by the pure dephasing with $1/f$ flux noise only. The large deviation of $\eta'$ from the pure dephasing limit when $\omega_{01}$ tuned far away from the sweet-point shows there exist additional dephasing sources, e.g. the fluctuations of $\Gamma_1$. Right $y$-axis: derived equivalent dephasing rate $\gamma_{eqv}$ over the same frequency range. The dashed-dotted line represents a pure dephasing rate due to the $1/f$ flux noise with spectral density $S_\Phi(\omega) = \frac{A_\Phi}{f}$, where $A_\Phi^{1/2} = 1.5\times10^{-6} \Phi_0$.}
\end{figure}

Furthermore, we also derive the emission efficiency $\eta'$ over a wide frequency range for different $\omega_{01}$, as shown with the left $y$-axis in Fig.~\ref{fig3}, by measuring the reflection $r_e$ with varied probe power, similar to the one in Fig.~\ref{fig2}(b), at different flux bias points. In the figure, one can recognize the suppression of emission efficiency due to the coupling to TLS defects, see the sharp abnormal drop near the flux bias sweet-point. We also estimate the effect of low frequency $1/f$ flux noise when $\omega_{01}$ is tuned far away from the sweet-point, which usually leads to a larger dephasing rate. The total dephasing rate is $\Gamma_2 = \Gamma_1/2 + \gamma$ and therefore we define $\gamma_{eqv}=(1-\eta')\Gamma_2=\frac{\Gamma_1^c + \Gamma_1^n}{2}+\gamma$. The derived $\gamma_{eqv}$ is shown with the right $y$-axis in Fig.~\ref{fig3}. 

The pure dephasing rate $\gamma$ due to $1/f$ flux noise with the flux noise spectral density 
$S_{\Phi}(\omega)=\frac{A_{\Phi}}{f}$ is $\gamma = \zeta \frac{A_{\Phi}^{1/2}}{\hbar}|\frac{\partial E_{01}}{\partial\Phi}|$ \cite{astafiev_noise_2004,ithier_decoherence_2005,shnirman_noise_2002}. The $\zeta$-factor is a result of the integration of $1/f$ noise, which is weakly dependent on the integration limits and in our case can be approximately taken as 3.5. A dashed-dotted line with the right $y$-axis in Fig.~\ref{fig3} shows calculated $\gamma$ of the $1/f$ noise with $A_{\Phi}^{1/2}=1.5\times10^{-6}~\Phi_0$, which is typical for the properly shielded devices \cite{yoshihara_decoherence_2006,kakuyanagi_dephasing_2007,bialczak_phasequbit_2007,hutchings_tunable_2017}. Accordingly, we also calculate the emission efficiency $\eta'$ limited only by the pure dephasing $\gamma$ with the same $1/f$ noise level, which is shown as a dashed line with the left $y$-axis in Fig.~\ref{fig3}. We see that $\eta'$ and $\gamma_{eqv}$ deviates more from the pure dephasing limit lines when tuned far away from the sweet-point and it can be caused by fluctuations of $\Gamma_1$ due to other reasons. 

Apart from the TLS defects, we achieve emission efficiency $\eta'\ge90\%$ over 1~GHz frequency range. Further improvement of the device to approach the pure dephasing limit could be better shielding in both low and high frequency regime or an improved design with less sensitivity to the flux-dependent dephasing, such as an asymmetrical design of the dc-SQUID \cite{hutchings_tunable_2017}. Additional resource for improving the state preparation efficiency can be the controllable coupling shown in Ref.~\cite{forn-diaz_-demand_2017}. However, it will require careful control of the coupling circuit to prevent further decrease of the coupling efficiency.    

\section{\label{sec:rab}Dynamics of emission field}
To further study the emission dynamics we utilize an approach similar to the one used in Ref. \cite{bozyigit_antibunching_2011}. We measure the quadrature amplitude and power of spontaneous emission from a pulse-driven qubit, namely $\langle a\rangle$ and $\langle a^{\dagger}a\rangle$ of the radiation mode. By using two independent detection channels $a$ \& $b$ \cite{menzel_dual-path_2010}, we can average out the uncorrelated noise in each channel, which makes the measurements of microwave single photons, using linear detectors \cite{da_silva_schemes_2010} and long-time averages, more efficient when compared with using just a single detection channel. 
 
\begin{figure*}
	\includegraphics[width=0.7\textwidth]{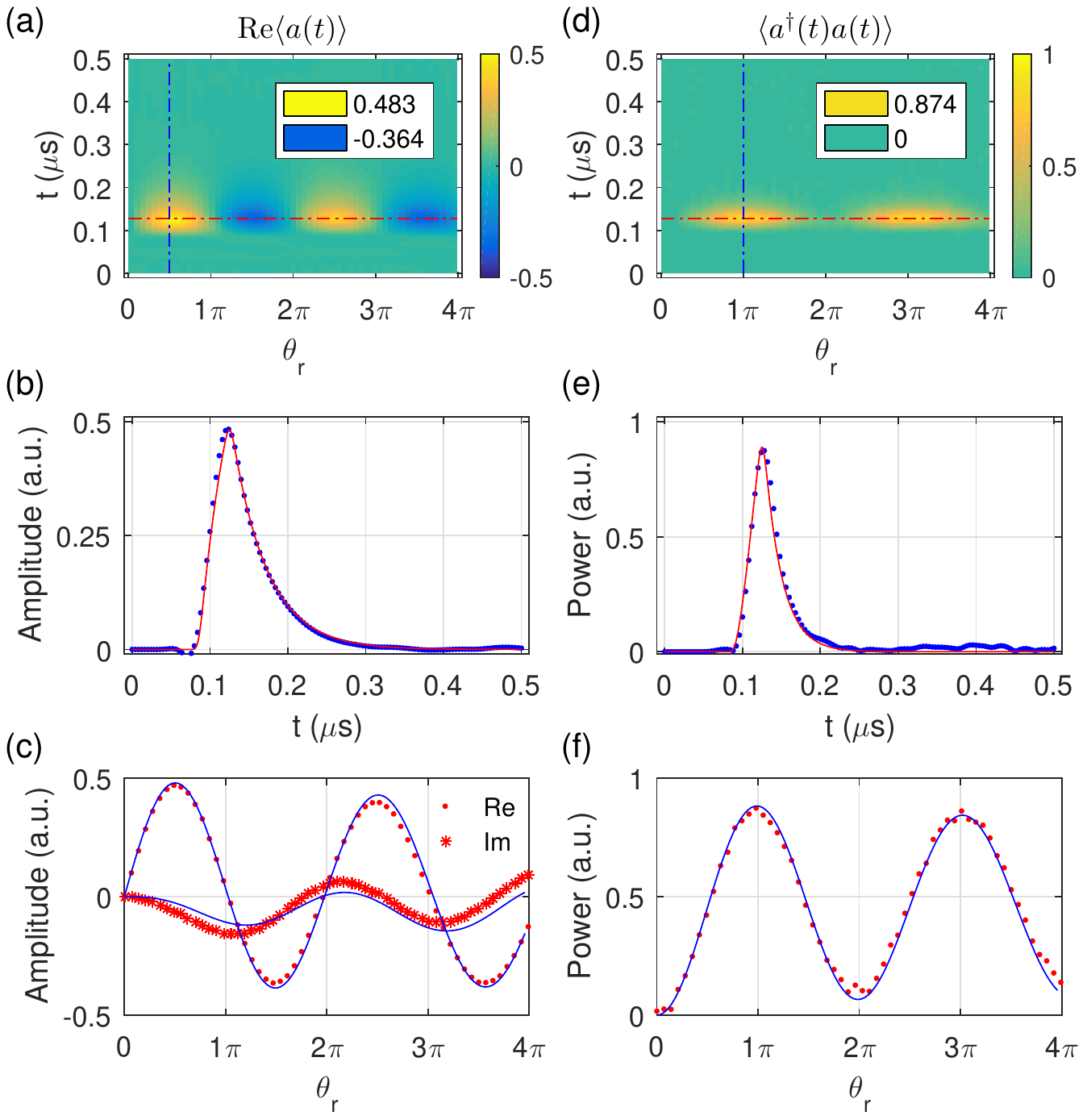}
	\caption{\label{fig4} (a) Time dependence of measured quadrature amplitude (real part) of emission field at a single channel of a beam splitter versus Rabi angle $\theta_r$. The legend shows colors for the maximum and minimum values. Global phase has been adjusted to minimize the imaginary part. (b) Single amplitude trace at $\theta_r=\pi/2$, corresponding to the state $(|0\rangle+|1\rangle)/\sqrt{2}$ (blue dashed-dotted line in (a)). (c) The dependence of maximum quadrature amplitude, including both real and imaginary parts, on $\theta_r$ at time $t_m$ indicated by the red dashed-dotted line in (a). (d) Measured time dependence of cross-power between two channels for the same Rabi angle as (a). (e) Single power trace at $\theta_r=\pi$ corresponding to state $|1\rangle$ (blue dashed-dotted line in (d)). (f) The dependence of maximum cross-power on $\theta_r$ at time $t_m$ indicated by the red dashed-dotted line in (d). All the dots are experimental data, and the solid lines are simulation results using the model explained in Sec.~\ref{sec:sim} with a finite detection bandwidth of 25 MHz.}
\end{figure*}

We apply a truncated Gaussian pulse $A\exp(-t^2/{2\sigma^2})$ of $\sigma=2$ ns and controlled amplitude $A$ at the control line to coherently control the transmon qubit. We can prepare the qubit in the state $\cos({\theta_r}/2)|0\rangle+\sin({\theta_r}/2)|1\rangle$, where $\theta_r$ is an angle acquired in the Rabi oscillation process (Rabi angle). As shown in Fig.~\ref{fig4}(a, d), we can observe the full dynamics of time dependence of the emission quadrature amplitude $\langle{a}(t)\rangle\propto\langle{S_a}(t)\rangle$ and power $\langle{a^{\dagger}}(t){a}(t)\rangle\propto\langle{
	S_a^*}(t){S_b}(t)\rangle$, characterized by the Rabi angle $\theta_r$ with $5\times10^7$ ensemble averages. Instead of calculating the direct power $\langle {S_a^*}(t){S_a}(t)\rangle$ in a single channel, we calculate the cross-power $\langle {S_a^*}(t){S_b}(t)\rangle$ between two channels of the beam splitter \cite{da_silva_schemes_2010,bozyigit_antibunching_2011}, which can greatly suppress the uncorrelated noise in each channel and result in a much lower effective noise temperature ($\sim22$ mK) \cite{bozyigit_correlation_2011}. 

As expected, the quadrature amplitude $\langle{a(t)}\rangle$ shows $(\sin\theta_r)/2$ dependence (see Fig.~\ref{fig4}(c)), while power is $\langle{a^{\dagger}}(t)a(t)\rangle\propto\sin^2({\theta_r}/2)$ (see Fig.~\ref{fig4}(f)). When $\langle{a^{\dagger}}(t){a}(t)\rangle$ is maximal ($\theta_r=\pi$), which corresponds to the excited state $|1\rangle$ in the qubit and single-photon emission, the quadrature amplitude $\langle{a(t)}\rangle$ instead becomes minimal. This shows that the single-photon emission is with determined photon number but uncertain phase. With short pulses, the excitation of higher energy level $|2\rangle$ may also be possible, due to the weak anharmonicity in the transmon qubit \cite{koch_charge-insensitive_2007,motzoi_simple_2009}. In Fig.~\ref{fig4}(c), we also show the imaginary part of $\langle{a(t)}\rangle$ and it cannot be calibrated to zero by adjusting the global phase \cite{bozyigit_antibunching_2011}. 
The master equation calculations (see details in Sec.~\ref{sec:sim}) accounting the energy level $|2\rangle$ well reproduce the experiment. The population of energy level $|2\rangle$ is calculated to be about $0.003$ when $\theta_r=\pi$. The decoherence decreases fidelity of Rabi oscillations in both quadratures of amplitudes and power, when the driving amplitude is increased. Due to the limited length of the state preparation pulse by the anharmonicity of transmon qubit ($\sim415$~MHz) and the bandwidth of 1~GS/s AWG ($\sim400$~MHz), here the efficiency to prepare state $|1\rangle$ is $\sim0.87$, very close to the simulated value $\sim0.88$ and the total efficiency to generate single photon is estimated to be $\sim0.86$. As discussed in Sec.~\ref{sec:eff}, the state preparation efficiency can be improved with the better intrinsic coherence time and the smaller coupling to emission line. 

To obtain the dephasing rate $\Gamma_2$ \cite{abdumalikov_dynamics_2011}, we exponentially fit the decay envelope of $\langle{a}(t)\rangle$ when transmon is prepared with $\theta_r=\pi/2$, as shown in Fig.~\ref{fig4}(b), which gives us $\Gamma_2/2\pi=3.54 \pm 0.05$ MHz. Similarly, we extract the relaxation rate $\Gamma_1/2\pi=7.02\pm0.25$ MHz, which matches the linewidth measured in spectrum, by preparing the transmon at $|1\rangle$ with $\theta_r=\pi$ and exponentially fitting the decay envelope of $\langle{a^{\dagger}}(t)a(t)\rangle$, see Fig.~\ref{fig4}(e). 

In Fig.~\ref{fig4}(b), (c), (e) \& (f), the experimental data are shown in dots and all solid lines are numerical simulation results using the Lindblad master equation under pulse drive with only two fitting parameters $\Gamma_1$ and $\Gamma_2$. The simulations here also account the limited detection bandwidth of 25 MHz in experiment and the theoretical temporal shapes reproduce the experimental results well, see Fig.~\ref{fig4}(b) \& (e). However the limited detection bandwidth of 25 MHz does not change the decay shapes.
 
\section{\label{sec:sim}Theoretical modeling}
To theoretically model the dynamics of our system, we take the Lindblad master equation approach. We consider the system as a pulse driven, ladder-type three-level ($|0\rangle, |1\rangle \& |2\rangle$) system coupled to a transmission line. 

The time-dependent system Hamiltonian in a rotating-frame at drive frequency $\omega_d$ and after rotating-wave approximation ($\hbar\equiv1$):  
\begin{equation}
H(t)=\left(
\begin{array}{ccc}
0&\frac{\Omega(t)}{2}&0\\
\frac{\Omega(t)}{2}&\delta&\frac{\lambda\Omega(t)}{2}\\
0&\frac{\lambda\Omega(t)}{2}&\alpha+2\delta
\end{array}\right)
\label{equ:3}
\end{equation} 
where $\delta=\omega_{01}-\omega_d$, $\alpha=\omega_{12}-\omega_{01}$, $\lambda=\sqrt{2}$ \cite{motzoi_simple_2009,koch_charge-insensitive_2007} and $\Omega(t)$ is a time-dependent drive strength with the Gaussian shape of $\Omega(t) = \Omega_0\exp(-t^2/{2\sigma^2})$.
 
The time-dependent Lindblad master equation for the density matrix $\rho$ is
\begin{eqnarray}
\dot{\rho}(t)=&-&\frac{i}{\hbar}[H(t),\rho(t)]\nonumber\\
+\sum_n\frac{1}{2}[2C_n\rho(t)C_n^+&-&\rho(t)C_n^+C_n-C_n^+C_n\rho(t)]~,
\label{equ:4}
\end{eqnarray}
where $C_n=\sqrt{\gamma_n}A_n$ are collapse operators, and $A_n$ are the operators through which the system couples to environment modes. 

Specifically, here we take:
\begin{eqnarray}
C_1=\sqrt{\Gamma_1}\sigma_{01}~&,&~C_2=\sqrt{\frac{\gamma}{2}}\sigma_{11}~,\\
C_3=\sqrt{2\Gamma_1}\sigma_{12}~&,&~C_4=\sqrt{\frac{\gamma}{2}}\sigma_{22}~,\nonumber
\label{equ:5}
\end{eqnarray}
where $\sigma_{jk}=|j\rangle\langle k|$ with $\{|j\rangle,|k\rangle\}=\{|0\rangle,|1\rangle,|2\rangle\}$.

Now we have only modeled our system using internal system operators which are not straightforwardly related to the photon emission. Then input-output theory \cite{walls2008quantum} provides us a direct connect between the internal system operator $\sigma_{01}$ and the external radiation mode operator $a$ by $a=\sqrt{\Gamma_1}\sigma_{01}$.
 
For dynamics of emission field in both quadrature amplitude and power, we have 
\begin{eqnarray}
\langle a(t)\rangle&=&\sqrt{\Gamma_1}\langle\sigma_{01}(t)\rangle~,\label{eqn:6}\\
\langle a^\dagger(t)a(t)\rangle&=&\Gamma_1\langle\sigma_{10}(t)\sigma_{01}(t)\rangle~.\label{eqn:7}
\end{eqnarray}

For correlation functions, there are
\begin{eqnarray}
G^{(1)}(t,\tau)&=&\langle a^\dagger(t)a(t+\tau)\rangle\nonumber\\
&=&\Gamma_1\langle\sigma_{10}(t)\sigma_{01}(t+\tau)\rangle~,\label{eqn:8}\\
G^{(2)}(t,\tau)&=&\langle a^\dagger(t)a^\dagger(t+\tau)a(t+\tau)a(t)\rangle\nonumber\\
&=&\Gamma_1^2\langle\sigma_{10}(t)\sigma_{10}(t+\tau)\sigma_{01}(t+\tau)\sigma_{01}(t)\rangle~.\label{eqn:9}
\end{eqnarray}  

We numerically solve the Lindblad master equation described above with real experimental parameters using QuTip \cite{johansson_qutip:_2012,johansson_qutip_2013} to simulate the time-evolution of our system. The anharmonicity of transmon $\alpha/2\pi\approx-415$ MHz is measured with two-tone spectrum (not shown). $\gamma=\Gamma_2-{\Gamma_1}/2$ with the fitted $\Gamma_1$ and $\Gamma_2$ in Sec.~\ref{sec:rab}. The driving strength $\Omega(t)$ has a similar truncated Gaussian pulse envelope of $\sigma=2$ ns and controlled amplitude $\Omega_0$. We found a good match between theoretical simulations (solid lines) and experimental data (dots) in Fig.~\ref{fig4}(c)\&(f). Note here, if with the infinite detection bandwidth in simulation, the theoretical simulations can show fast dynamics inside the rising envelope of temporal pulse shape when $\theta_r\ge\pi$, which will help us to understand the physics of system better \cite{fischer_pulsed_2018}. 

For correlation functions, we can also quickly calculate the two-time correlation of different operators using QuTip built-in functions which uses quantum regression theorem \cite{gardiner2004quantum}. The solid lines showed in Fig.~\ref{fig5}(a)-(c) are numerical calculation results with consideration of finite detection bandwidth. We found a good agreement between simulations and experiments. 
         
\section{\label{sec:corr}Correlation function measurement}
Furthermore, we measure the correlation functions of the emitted photons using linear detectors \cite{da_silva_schemes_2010} with HBT setup, see Fig.~\ref{fig1}(b), to confirm the single-photon emission of our source. Because of the near unit collecting efficiency of emitted photons and more efficient signal processing technique used here, we see also a much better SNR in the second-order correlation function measurement than our previous work \cite{peng_tuneable_2016}. 

We generate a train of 16 single-photon pulses ($\theta_r=\pi$) with a separation of $t_p=512 ~\text{ns}~(t_p\gg1/\Gamma_1)$ between two adjacent pulses. This ensures that the transmon qubit always returns to its ground state before being excited again. The emitted photons are split into two channels by the hybrid coupler and then are amplified at both 4K and RT stages. Next, the signals from two channels are down-converted to IF signals at 25~MHz and filtered by 48~MHz low-pass filter. Finally the IF signals are digitized by two ADCs and further processed by the CPU with a GPU-enhanced signal processing technique to calculate the correlations between two quadrature amplitudes $S_a(t)$ and $S_b(t)$. 

For calculation of correlation functions, we follow the same way as in Ref.~\cite{bozyigit_antibunching_2011,da_silva_schemes_2010}:
\begin{equation}
\Gamma^{(1)}(\tau)=\int\langle S_a^*(t)S_b(t+\tau)\rangle dt~,
\label{eqn:10}
\end{equation}
which measures the first-order cross-correlation of signal $S_a(t)$ and $S_b(t)$.
\begin{equation}
\Gamma^{(2)}(\tau)=\int\langle S_a^*(t)S_a^*(t+\tau)S_b(t+\tau)S_b(t)\rangle dt~,
\label{eqn:11}
\end{equation}
which measures the quasi-auto correlation of cross-power $S_a^*(t)S_b(t)$ and serves as a measurement of second-order correlation function only in the HBT-like setup with an idle input port in vacuum. 

To remove the correlated noise background ($\Gamma_{bg}^{(1)}(\tau)$ and $\Gamma_{bg}^{(2)}(\tau)$), each signal trace is immediately followed by a trace of noise background, when the photon source is not excited. Signal traces and noise background traces are calculated and averaged in the same way to get the correlation function. Then, we can obtain the correlation function of emitted photons by
\begin{eqnarray}
G^{(1)}(\tau)&\propto&\Gamma^{(1)}(\tau)-\Gamma_{bg}^{(1)}(\tau)~,\label{eqn:12}\\
G^{(2)}(\tau)&\propto&\Gamma^{(2)}(\tau)-\Gamma_{bg}^{(2)}(\tau)~.\label{eqn:13}
\end{eqnarray}

\begin{figure*}
	\includegraphics[width=0.7\textwidth]{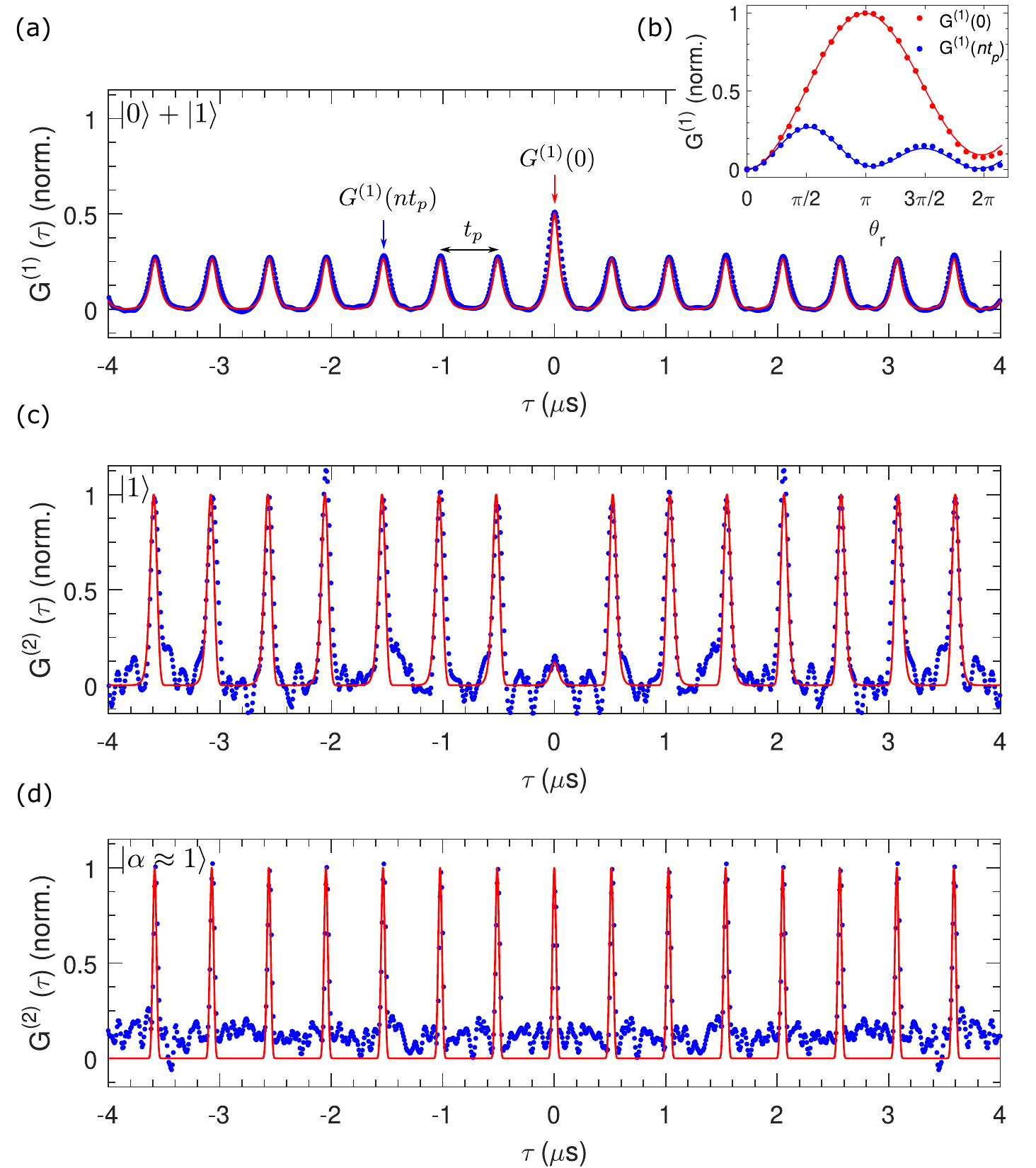}
	\caption{\label{fig5} (a) Time dependence of the first-order correlation function $G^{(1)}(\tau)$ for $(|0\rangle+|1\rangle)/\sqrt{2}$. (b) The dependence of a center peak $G^{(1)}(0)$ and a side-peak $G^{(1)}(nt_p)$ on the prepared Rabi angle $\theta_r$. (c) The measured second-order correlation function $G^{(2)}(\tau)$ for the state $|1\rangle$. (d) The measured second-order correlation function $G^{(2)}(\tau)$ for the coherent state $|\alpha\approx1\rangle$. All the dots are experimental data, and solid lines are theoretical calculations including the effect of limited detection bandwidth in experiment.} 
\end{figure*}

For $G^{(1)}(\tau)$, the trace (dots) shown in Fig.~\ref{fig5}(a) is averaged by $5\times10^7$ trains of 16 photons with specific prepared state. Fig.~\ref{fig5}(b) shows the dependence of $G^{(1)}(0)$ and $G^{(1)}(nt_p)$ on the Rabi angle $\theta_r$. The center peak $G^{(1)}(0)\propto\langle a^{\dagger}a\rangle$ measures the average emitted number of photons. As photons generated in different pulses are not correlated ($t_p\gg1/\Gamma_1$), $G^{(1)}(nt_p)\propto\langle a^{\dagger}\rangle\langle a\rangle$. The solid lines are simulation results using master equation (see Sec.~\ref{sec:sim}) with a finite detection bandwidth of 20 MHz here. The damping in oscillation is mainly due to the decoherence. We see an excellent agreement between our theoretical calculations and experiment results.

For $G^{(2)}(\tau)$, the measurement in the microwave regime is technically difficult due to the very poor SNR, which needs extremely large averaging ($>10^9$) to achieve a reasonable confidence in experimental data. Instead of using FPGA to realize real-time signal processing \cite{bozyigit_antibunching_2011,lang_correlations_2013}, here we take an intermediate approach using 1792 CUDA cores in GPU to realize parallel signal processing, which can speed-up by 4x when compared to solely using one Xeon CPU of 6 cores. This approach demonstrated here is not as efficient as FPGA but easier to achieve and also has higher flexibility. The trace (dots) in Fig.~\ref{fig5}(c) is averaged $4.8\times10^9$ times within 34 hours, corresponding to $\sim 36$ TB data processed in total. The data has been normalized to the average peak height of $G^{(2)}(nt_p)$ for state $|1\rangle$ ($\theta_r=\pi$) \cite{bozyigit_antibunching_2011}. We see a strongly suppressed center peak $G^{(2)}(0)\approx0.15\ll1$ which is limited by $\Gamma_1T_{FWHM}$, where $T_{FWHM}$ is the length of state preparation pulse \cite{fischer_dynamical_2016}. This result shows a clear evidence of the single-photon emission from our source. Here the detection bandwidth is set at 12.5 MHz to suppress the noise outside of signal bandwidth $\sim7$ MHz. The solid line is a result of the master equation simulation with the experimental detection bandwidth and $G^{(2)}(0)\approx0.1$ is very close to the experimental value.

To further confirm our experimental results, we carry out another experiment for comparison. We direct generate a short coherent Gaussian pulse with $\sigma=8$ ns and the power is calibrated to make sure that the average photon number inside the pulse is approximately one. Then we measure the second-order correlation function $G^{(2)}(\tau)$ of this coherent state $|\alpha\approx1\rangle$ with $4\times10^9$ averages. As expected, we observe $G^{(2)}(0)\approx0.9$ at $\tau=0$, which is in significant contrast to the Fock state $|1\rangle$. The red line is the theoretical calculation with a real pulse temporal shape and a finite detection bandwidth. The noise background is slightly higher than Fig.~\ref{fig5}(c) because of the broadband noise existing in this particular experimental setup with a transmission line.
  
Note here, we demonstrate the antibunching of photons emitted from our source. In further experiments, we can also demonstrate indistinguishability of photons, using Hong-Ou-Mandel (HOM) effect \cite{hong_measurement_1987,santori_indistinguishable_2002}. The demonstration of the HOM effect from the same single-photon source in the microwave regime is challenging because of the difficulty to introduce the large time-delay (hundreds of ns) at cryogenic temperatures. Due to the strong coupling to the 1D continuum, we should have negligible pure dephasing ($\gamma \approx 0$) \cite{astafiev_resonance_2010} and $2\Gamma_2/\Gamma_1=1$. If this condition is satisfied, the perfect two-photon interference can be expected \cite{santori_indistinguishable_2002,bylander_interference_2003}. Here in our source, $2\Gamma_2/\Gamma_1\approx1$ which indicates good two-photon interference. Thus, the higher emission efficiency $\eta'$ also means the better indistinguishability of emitted photons from the source, which is crucial for practical applications.
     
\section{\label{sec:cons}Conclusion}
In conclusion, we have demonstrated a highly efficient tunable microwave single-photon source based on a transmon qubit with the intrinsic emission efficiency of $\sim0.99\pm0.01$. Considering the state preparation efficiency of $\sim0.87$, the total single-photon generation efficiency is $\sim0.86\pm0.01$. The single-photon emission of our source is also confirmed by correlation functions measurement using a GPU-enhanced signal processing technique.
In this work, we have substantially improved the performance of a tunable microwave single-photon source. The achieved above 98$\%$ emission efficiency is a result of the negligible pure dephasing rate, which is a strong evidence that the emitted photons are indistinguishable. Further analysis also shows that the state preparation efficiency can be further improved with better intrinsic coherence time (e.g. $T_1^n\sim 20~\mu$s) and smaller coupling to emission line (e.g. $T_1^e\sim0.2~\mu$s). Our result shows that such a tunable microwave single-photon source using waveguide scheme can be good for various practical applications in quantum communication, simulation and information processing in the microwave regime.

\begin{acknowledgments}
Y. Zhou would like to thank D.K. Zhang, R. Wang, N. Lambert and A. Miranowicz for valuable discussion, H. Mukai for help in experimental setup and K. Kusuyama for help in fabrication. This work was supported by CREST, JST. (Grant No. JPMJCR1676), the New Energy and Industrial Technology Development Organization (NEDO), and ImPACT Program of Council for Science, Technology and Innovation (Cabinet Office, Government of Japan). Z.H.P. is supported by NSFC under Grant No. 61833010 and Hunan Province Science and Technology Innovation Platform and Talent Plan (Excellent Talent Award) under grant No.2017XK2021. O.V.A. is supported by Russian Science Foundation (grant N 16-12-00070).
\end{acknowledgments}

\bibliography{single_photon_source_rev1128.bib}

\end{document}